\documentclass{cjpsuppl}% for LaTeX 2e
\usepackage{graphicx} % or \usepackage{epsfig }
\begin{document}
\title{Hard Electroproduction of Hybrid Mesons}
\authori{I.V.~Anikin}
\addressi{Bogoliubov Laboratory of Theoretical Physics, JINR, 141980 Dubna,
Russia and \\ 
LPT Universit{\'e} Paris-Sud, 91405-Orsay,
France }
\authorii{B.~Pire}    \addressii{CPHT, {\'E}cole
Polytechnique, 91128 Palaiseau, France}
\authoriii{ L.~Szymanowski}   \addressiii{Soltan Institute for Nuclear 
 Studies, Warsaw, Poland and \\
 Inst. de Physique,
Univ. de Li{\`e}ge, B-4000 Li{\`e}ge, Belgium}
\authoriv{O.V.~Teryaev}    \addressiv{Bogoliubov Laboratory of Theoretical Physics, JINR, 141980 Dubna,
Russia }
\authorv{S.~Wallon} \addressv{LPT, Universit{\'e} Paris-Sud, 91405-Orsay, France}
\authorvi{}    \addressvi{}
\headtitle{ Hard Electroproduction \ldots}
\headauthor{Anikin}
\lastevenhead{Anikin: Hard Electroproduction \ldots}
%%%%%%%%%%%%%% Pro editory supplementu: %%%%%%%%%%%%%%%
\refnum{A229-A234}
\daterec{} %Date of send of contribution
\suppl{A}  \year{2005} \setcounter{page}{1}
 %\firstpage{1}
 %\lastpage{000}
 %\makefirsttitle
 %%%%%%%%%%%%%%%%%%%%%%%%%%%%%%%%%%%%%%%%%%%%%%
 \maketitle

\begin{abstract}
We estimate the sizeable cross section for deep exclusive electroproduction of an exotic
$J^{PC}=1^{-+}$ hybrid meson in the Bjorken regime.  The production amplitude scales like
the one for usual meson electroproduction, {\it i.e.} as
$1/Q^2$. This is due to the non-vanishing  leading twist distribution
amplitude for the hybrid meson, which may be normalized thanks to its relation to the
energy momentum tensor and to the QCD sum rules technique. The hard amplitude is
considered up to next--to--leading order in $\alpha_{S}$ and we explore the consequences
of fixing the renormalization scale ambiguity through the BLM procedure.
\end{abstract}

\section{ Introduction}

Within quantum chromodynamics, hadrons are described in terms of quarks, anti-quarks and
gluons. The usual, well-known, mesons are supposed to contain quarks and
anti-quarks as valence
degrees of freedom while gluons play the role of carrier of interaction,
{\it i.e.} they remain hidden in a background.
On the other hand, QCD does not prohibit the existence of the explicit gluonic degree of freedom
in the form of a vibrating flux tube, for instance. The states where the $q\bar q g$ and
$gg$ configurations are dominating, hybrids and glueballs,
are of fundamental importance to understand the dynamics of quark confinement and the
nonperturbative sector of quantum chromodynamics \cite{Close},\cite{Jaffe}.

The study of these hadrons outside the constituent quark models, namely exotic hybrids, is the main 
reason of the present paper. We investigate how hybrid mesons with $J^{PC}=1^{-+}$
may be studied through the so-called hard reactions.
We focus on deep exclusive meson electroproduction
(see, for instance \cite{Goe}) which is well described in the framework of the collinear
approximation where generalized parton distributions (GPDs) \cite{Muller:1994fv} and distribution
amplitudes \cite{ERBL} describe the nonperturbative parts of a factorized amplitude 
\cite{Collins:1997fb}.

\section{Hybrid meson production amplitude}

We propose to study the exotic hybrid meson by means of its deep exclusive
electroproduction, {\it i.e.}
\begin{eqnarray}
\label{pr}
e(k_1)\, + \, N(p_1)\,\to\,e(k_2)\,+ H(p)\,+\,N(p_2),
\end{eqnarray}
where we will concentrate on the subprocess:
\begin{eqnarray}
\label{pr2}
\gamma^*_L(q)\, + \, N(p_1)\,\to\, H_L(p)\,+\,N(p_2)
\end{eqnarray}
when the baryon is scattered at small angle.
This process is a hard exclusive reaction due to the transferred momentum $Q^2$
is  large ( Bjorken regime).
Within this regime, a factorization theorem is valid,  at
the leading twist level, which claims that a partonic subprocess part described in perturbative
QCD (pQCD)
can be detached from universal soft parts, which are generalized parton distributions and meson
distribution amplitudes.
Below we will analyze in more details how this factorization theorem applies to the  process
under study.

Let us fix the kinematics of the deep electroproduction process. We are interested in the
scaling regime where the virtuality of the
photon $Q^2=-q^2$ is large and scales with the energy of the process.
We denote by $p_1$ ($p_2$) the momentum of the incoming (outgoing) nucleon,
while $p$ is the momentum of the longitudinally polarized hybrid meson of mass $M_{H}$.
We construct the average momentum
$\overline{P}$ and transferred momentum $\Delta$:
\begin{eqnarray}
\label{PDel}
&&\overline{P}=\frac{p_2+p_1}{2}, \quad \Delta=p_2-p_1, \quad \Delta^2=t.
\end{eqnarray}
With two light-cone vectors: $n^{*\,2}= n^2=0, \, n^*\cdot n=1$,
the Sudakov decompositions for  all the
relevant momenta take the form:
\begin{eqnarray}
\label{Sd}
&&\Delta_\mu=-2\xi n^*_\mu +\xi \overline{M}^2 n_\mu + \Delta^T_\mu,
\quad \Delta^T\cdot n=\Delta^T\cdot n^*=0,
\nonumber\\
&&\overline{P}_\mu=n^*_\mu+\frac{\overline{M}^2}{2}n_\mu , \quad \overline{P}^2=
\overline{M}^2,
\quad \xi\leq \frac{\sqrt{-\Delta^2}}{2\overline{M}}\leq 1,
\nonumber\\
&&q_\mu=-2\tilde\xi n^*_\mu +\frac{Q^2}{4\tilde\xi}n_\mu, 
\nonumber\\
&&p_\mu=q_\mu-\Delta_\mu=2(\xi-\tilde\xi) n^*_\mu+
\biggl( \frac{Q^2}{4\tilde\xi}-\xi\overline{M}^2 \biggr)n_\mu -\Delta^T_\mu.
\end{eqnarray}
Here, the parameters $\xi$ and $\tilde\xi$ are related by
\begin{eqnarray}
M_{H}^2=4(\xi-\tilde\xi)
\biggl(\frac{Q^2}{4\tilde\xi}-\xi\overline{M}^2 \biggr)+\Delta_T^2.
\end{eqnarray}
The leading order amplitude for the process (\ref{pr}) is
\begin{eqnarray}
\label{amp02}
{\cal A}_{(q)}= \frac{e\pi\alpha_s f_{H} C_F}{\sqrt{2}N_c Q}
\biggl[ e_u {\cal H}_{uu}^- -e_d {\cal H}_{dd}^-\biggr] {\cal V}^{(H,\,-)},
\end{eqnarray}
where
\begin{eqnarray}
\label{softin1}
&&{\cal H}_{ff}^\pm =
\int\limits_{-1}^{1}dx \biggl[
\overline{U}(p_2)\hat n U(p_1) H_{ff^{\prime}}(x) +
\overline{U}(p_2)\frac{i\sigma_{\mu\alpha} n^{\mu}\Delta^{\alpha}}{2M}U(p_1)
E_{ff^{\prime}}(x) \biggr]
\nonumber \\
&&
\biggl[
\frac{1}{x+\xi-i\epsilon}\pm\frac{1}{x-\xi+i\epsilon}\biggr],
\quad
%\nonumber\\
%\label{softin2}
{\cal V}^{(M,\,\pm)}=
\int\limits_{0}^{1} dy \phi^{M}(y)\biggl[
\frac{1}{y}\pm\frac{1}{1-y}
\biggr].
\end{eqnarray}
Here, functions $H$ and $E$ are standard leading twist GPD's and their properties are
fairly well-known.
In (\ref{softin1}), we include the definition of ${\cal H}_{ff}^+$ and 
${\cal V}^{(M,\,+)}$ which will be useful for the comparison with the $\rho$ meson 
case. The hybrid meson distribution amplitude is a new object
and we will carefully study it in the next subsection.
We will now consider the properties of the hybrid meson
distribution amplitude (see also \cite{APSTW1}--\cite{APSTW3}).
The Fourier transform of the hybrid meson --to--vacuum matrix element of
the bilocal vector quark operator may be written as
\begin{eqnarray}
\langle H_L(p,0)| \bar \psi(-z/2)\gamma_\mu 
\psi(z/2) | 0 \rangle =
i f_H M_H e^{(0)}_{L\,\mu}
\int\limits_0^1 dy e^{i(\bar y - y)p\cdot z/2} \phi^{H}_L(y)
\label{hmeW}, 
\end{eqnarray}
where $e^{(0)}_{L\,\mu}=(e^{(0)}\cdot z)/(p\cdot z) p_\mu$
and $\bar y=1-y$ and $H$ denotes the isovector triplet of hybrid mesons;
$f_H$ denotes a dimensionful coupling constant of the
hybrid meson, so that $\phi^H$ is dimensionless. 

In (\ref{hmeW}),
we imply the path-ordered gluonic exponential along the straight line connecting
the initial and final points $[z_1;z_2]$ which provides the gauge invariance
for bilocal operator and
equals unity in a light-like (axial) gauge. For  simplicity of notation we shall omit the index
$L$ from the  hybrid meson distribution amplitude.

Although exotic quantum numbers like $J^{PC} = 1^{-+}$ are forbidden in the quark model,
it does not prevent the
leading twist correlation function from being non zero. The basis of the argument is that the
non-locality of the quark correlator opens the possibility of getting such a hybrid
state, because of dynamical gluonic degrees of freedom arising from the Wilson line 
(more details can be found in \cite{APSTW2}).

\section{Cross-sections for hybrid meson electroproduction}

The unpolarized cross section corresponding to the reaction  (\ref{pr2})
is defined by
\footnote{The flux factor is chosen as in \cite{Vand99}.}
\begin{eqnarray}
\label{xsecL}
\frac{d\sigma_L}{d\hat t}=\frac{1}{16\pi(\hat s-m_N^2)\lambda(\hat s,-Q^2,m_N^2)}
\frac{1}{2}\sum_{pol.} |{\cal A}_{(q)}|^2,
\end{eqnarray}
where the amplitude ${\cal A}_{(q)}$ is determined by (\ref{amp02});
$\hat s$, $\hat t$  are the usual Mandelstam variables
 and $m_N$ is the nucleon mass.
The function $\lambda$ is standardly defined by
\begin{eqnarray}
\label{lambda}
\lambda^2(x,y,z)=x^2+y^2+z^2-2xy-2xz-2yz.
\end{eqnarray}
To calculate the cross section (\ref{xsecL}), we need to model the corresponding GPD's.
We apply the Radyushkin model \cite{Rad} where the function $H$, see (\ref{amp02}),
is expressed with the help of double distributions $F^q(x,y;t)$.
\begin{figure}
$$\rotatebox{270}{\includegraphics[width=6cm]{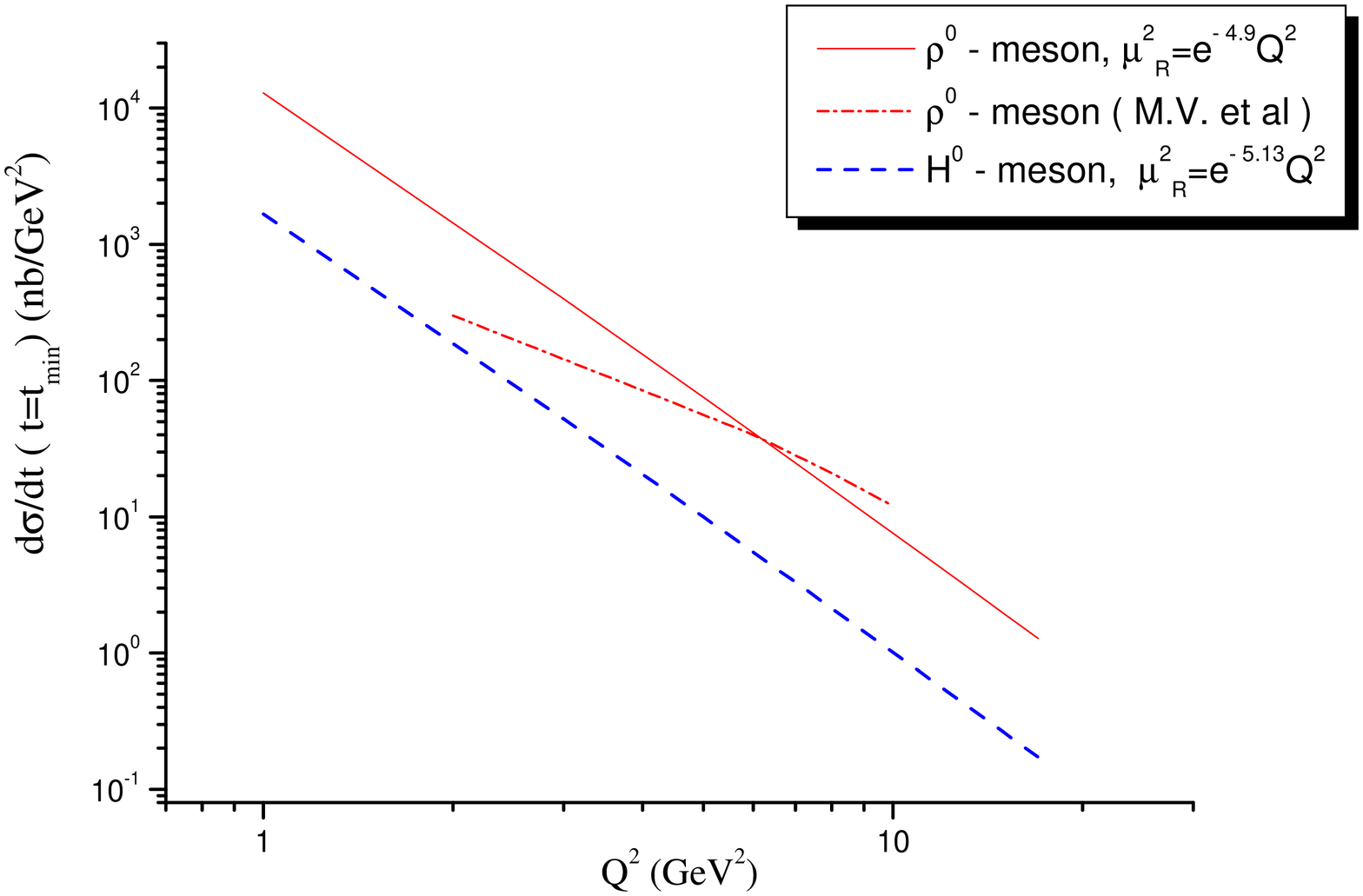}}$$
\caption{Differential cross-section for exotic hybrid meson
electroproduction (dashed line) with
$\mu_R^2 =e^{-5.13} Q^2$ compared with
the quark contribution to $\rho^0$
electroproduction (solid line) with $\mu_R^2 =e^{-4.9} Q^2$, as a function
of $Q^2,$ for $x_B\approx 0.33$. The dash-dotted line is the result of Vanderhaegen et al 
\cite{Vand99} for $\rho$ electroproduction.}
\label{xsec2}
\end{figure}
To get prediction for the cross sections we need to fix the renormalization 
scales. In order to estimate theoretical uncertainties of this procedure we 
fix the scale $\mu_R^2$ in two different ways: firstly, in the naive way, by assuming 
$\mu^2_R=Q^2$, and secondly, by applying the BLM prescription \cite{BLM}.
The BLM procedure, which is discussed in details in \cite{APSTW3},
leads to the following values of the renormalization scales:
\begin{eqnarray}
\label{BLMsc1}
&&\mu^2_R=e^{-4.9}Q^2, \quad \mbox{for\,\, $\rho$ \,\,meson}, 
\nonumber \\
&&\mu^2_R=e^{-5.13}Q^2, \quad \mbox{for\,\, $H$\,\, meson.} 
\end{eqnarray}
for the case $\xi=0.2$ (or $x_B\approx 0.33$).
These renormalization scales have rather small
magnitudes. This has a tendency to enlarge the cross sections but may endanger the validity
of the perturbative approach. However, it is possible that
the coupling constant $\alpha_S$ stays below unity and 
the perturbative theory does not suffer from the IR divergencies.
We will use the Shirkov and Solovtsov's ansatz \cite{Shirkov}
where the analytic running coupling constant takes the form:
\begin{eqnarray}
\label{asan}
\alpha_{S}^{an}(\mu_R^2) = \frac{4\pi}{\beta_0}
\biggl[
\frac{1}{{\rm ln} \mu_R^2/\Lambda_{QCD}^2}+\frac{\Lambda_{QCD}^2}{\Lambda_{QCD}^2-\mu^2_R}
\biggr].
\end{eqnarray}
Here $\Lambda_{QCD}$ is the standard scale parameter in QCD.
The second term in (\ref{asan}) assures the absence of a ghost pole at $\mu_R^2=\Lambda_{QCD}^2$
and has a nonperturbative source. Detailed discussion on this point may be found in
\cite{Bakulev} and references therein.

Recently, in \cite{Vand99} the role of power 
corrections due to the intrinsic transverse momentum
of partons (the kinematical higher twist) has been investigated.
In that approach the inclusion of the intrinsic transverse momentum
dependence results in a rather strong effect on the differential cross-section  
before the scaling regime is achieved. In \cite{Vand99}, the renormalization 
scale $\mu^2_R$ is defined by
the gluon virtuality so that the scale is a function of parton fractions
flowing into the corresponding gluon propagator. 
On Fig. \ref{xsec2}, we present our results for the differential cross section of the hybrid 
meson electroproduction compared to the $\rho$ meson electroproduction, using the BLM scales. 
We can see that the hybrid cross section is rather sizeable in comparison with 
the corresponding $\rho$ meson cross section.
One can see that in the region $Q^2\sim 5 - 10\, {\rm GeV}^2$ the size of
the $\rho$ meson cross section obtained  with the inclusion of  
transverse momentum effects 
is very close to the analogous cross section computed with the BLM scale and without
the intrinsic transverse momentum dependence. On the other hand, for higher values of $Q^2$
the leading order amplitude 
computed with the BLM scale fixing  is falling faster that the 
corresponding amplitude derived in Ref. \cite{Vand99}, whereas for smaller 
values of $Q^2$ it is larger than that prediction.

\section{Conclusion}

In conclusion, we have calculated 
the leading twist contribution to exotic hybrid meson with $J^{PC}=1^{-+}$ electroproduction
amplitude in the deep exclusive region.  The resulting order of magnitude
is somewhat smaller than the $\rho$ electroproduction but
similar to the $\pi$ electroproduction. The obtained cross section is
sizeable and should be measurable at dedicated experiments at JLab, Hermes or Compass.
We made a systematic comparison with the non-exotic vector meson production.
To take into account NLO corrections, the differential cross-sections for these processes have
been computed using the BLM prescription for the renormalization scale.
In the case of $\rho$ production, our estimate is not far from a previous one  which
took into account kinematical higher twist corrections.
In the region of  small $Q^2$  higher twist contributions should be carefully studied and
included. Note that they have already been considered in the case of deeply
virtual Compton scattering \cite{APT} where their presence was dictated by gauge invariance, and
for transversely polarized vector mesons \cite{AT} where the leading twist component vanishes.
We  leave this study for future works.

\bigskip

{\small We acknowledge useful discussions with A.~Bakulev, I.~Balitsky, V.~Braun, M.~Diehl,
G.~Korchemsky, C.~Michael, S.~Mikhailov and O.~Pene. 
This work is supported in part by INTAS (Project 00/587)
and RFBR (Grant 03-02-16816).
The work of B.~P., L.~Sz. and S. W. is partially supported by the French-Polish scientific 
agreement Polonium and the Joint Research Activity "Generalised Parton 
Distributions" of the european I3 program Hadronic Physics, contract 
RII3-CT-2004-506078. 
I.~V.~A. thanks  NATO for a Grant. L.~Sz. thanks CNRS for a Grant supported 
his visit to LPT in Orsay.
L.~Sz. is a Visiting Fellow of the Fonds National pour la Recherche
Scientifique (Belgium)}

 %\lastevenpage
 \end{document}